\documentstyle[12pt,a4wide]{article}

\catcode`@=12

\begin{document}
\hoffset=0.4cm
\voffset=-1truecm
\normalsize


\begin{titlepage}
\begin{flushright}
DFPD 94/TH/51
\end{flushright}
\vspace{12pt}
\centerline{\Large {\bf Implications of a Light Stop for the Spontanous
CP }}
\vskip 0.1 cm
\centerline{\Large {\bf Breaking at Finite Temperature }}
\vskip 0.1 cm
\centerline{\Large {\bf in a Nonminimal
Supersymmetric Model}}
\vspace{10pt}
\begin{center}
{\large\bf D. Comelli$^{a,b,}$\footnote{Email:
comelli@mvxtst.ts.infn.it}, M. Pietroni$^{c,d,}$\footnote{
Email:pietroni@vxdesy.desy.de}
and A. Riotto$^{d,e,}$\footnote{On leave of absence from
International School for Advanced Studies (ISAS), Trieste, Italy;
Email:riotto@tsmi19.sissa.it}}
\end{center}
\vskip 0.2 cm
{\footnotesize
\centerline{\it $^{(a)}$ Dipartimento di Fisica Teorica, Universita` di
Trieste, 34014, Italy}
\vskip 0.2 cm
\centerline{\it $^{(c)}$ DESY Theory Group, Deutsches
Elektronen-Synchroton}
\centerline{Notkerstrasse 85, 22603 Hamburg, Germany}
\vskip 0.2 cm
\centerline{\it $^{(d)}$ Istituto Nazionale di Fisica Nucleare,}
\centerline{\it Sezione di Trieste, 34014 Trieste, Italy}
\vskip 0.2 cm
\centerline{\it $^{(c)}$Istituto Nazionale di Fisica Nucleare,}
\centerline{\it Sezione di Padova, 35100 Padua, Italy}
\vskip 0.2 cm
\centerline{\it $^{(d)}$Instituto de Estructura de la Materia,CSIC,}
\centerline{\it Serrano, 123, E-20006 Madrid, Spain}}
\baselineskip=20pt
\vskip 0.5 cm
\centerline{\large\bf Abstract}
\vskip 0.2 cm
We study the implications of a light stop for the spontaneous CP breaking
at finite temperature in the Higgs sector of the Minimal Supersymmetric
Standard Model with a gauge singlet. Assuming CP conservation at zero
temperature, we show that the presence of a
large mixing between the left- and the right-handed stops can
trigger easily the spontaneous breaking of CP inside the bubbles
nucleated during the electroweak phase transition. This allows to avoid
the fine-tuning among the vacuum expectation values in the region
of interest for the generation of the baryon asymmetry, namely the
bubble walls, which has been recently analized.

\end{titlepage}
\baselineskip=20pt

The generation of the baryon asymmetry in the Universe (BAU) requires
three basic ingredients \cite{sak}: baryonic violating interactions,
departure from thermal equilibrium and C and CP violation. Anomalous
electroweak processes are known to provide the source of B violation
\cite{kuz}, whereas the departure from thermal equilibrium can occur if
the electroweak phase transition  (EWPT) is of the first order and
proceeds via bubble nucleation. These considerations make the
possibility of generating the BAU
during the electroweak phase transition very appealing
\cite{nelson}. As far as the CP violation is concerned,
it is
still an open question if the necessary amount of CP violation to
produce enough baryon asymmetry is present in the standard model
\cite{if}.

In the minimal supersymmetric extension of the standard model (MSSM)
\cite{haber} the requirement that explicit CP violating  phases
provide the necessary amount of CP violation necessary for the
generation of the BAU gives rise to additional strong constraints
on the parameter space of the model \cite{s}. Spontaneous CP breaking
(SCPB) in the Higgs sector
can be triggered by radiative corrections \cite{mae}
at zero temperature. Nevertheless, as predicted by Georgi and Pais \cite{gp},
it requires
the existence of a pseudoscalar Higgs boson with a mass of a few GeV
\cite{pom}, which has been ruled out by LEP \cite{lepp}. When
finite temperature effects are considered, even if SCPB in the MSSM
during the EWPT can occur in a wide region of the parameter space
\cite{cp,cpr} it requires as a general tendency small values
of the mass of the pseudoscalar Higgs boson, whereas recent results
on the phase transition in the MSSM
\cite{rec} seem to point towards the opposite direction in order not
to wash out the generated baryon asymmetry.

Very recently, the question of SCPB at finite temperature in the MSSM
with a gauge singlet, the so-called next-to-minimal supersymmetric
standard model (NMSSM) \cite{n}, has been addressed \cite{singlet} in
the case of negligible left-right mixing in the stop sector.
Assuming CP conservation at zero temperature and including the
contributions to the one-loop effective potential both of the standard
model particles and of the supersymmetric ones, it has been shown that
SCPB in the broken phase ({\it i.e.} inside the bubbles nucleated
during the EWPT) is
prevented by two main reasons: large plasma effects in the Higgs sector
and the great reduction at finite temperature of the large one-loop
corrections
coming from the top-stop sector at zero temperature when both
the stops present in the spectrum are
nearly degenerate ({\it i.e.} negligible left-right mixing) and with a
mass of order of the critical temperature \cite{singlet}. Even if the
SCPB can occur in the region of interest for the generation of the
baryon asymmetry, namely in the bubble walls, this requires a
fine-tuning among the values of the vacuum expectation values (VEV's)
inside the walls which makes the phenomenon not very appealing.

In this Letter we argue that, in the framework of NMSSM,
 SCPB can easily occur inside the
expanding bubbles thus avoiding any fine-tuning described in ref.
\cite{singlet}\footnote{We remind the reader that the VEV's
and the associated phases continously change from
inside to outside the bubbles where they are
vanishing so that nonvanishing phases will be present in the bubble
walls.} when a large left-right mixing in the stop sector is
considered. Indeed, in this case, a relatively light
stop $\tilde{t}_2$ appears in the spectrum and its
presence can have many implications, for
example as regards the $\rho$ parameter, $K^0-\overline{K}^0$ mixing,
the rare decay $b\rightarrow s+\gamma$ and the proton decay through
dimension-5 operators in grand unified extension of the supersymmetric
model \cite{light}. Very recently the OPAL Collaboration has excluded the
existence of a light stop with a mass below $\sim$ 45 GeV unless a
mixing angle $\theta_{\tilde{t}}$ between the left- and the right-handed
partners is in the range $0.85<\theta_{\tilde{t}}<1.15$ and the mass
difference between the $\tilde{t}_2$ and the lightest neutralino
is smaller than 5 GeV \cite{opal}. From now on we shall take the
conservative bound $m_{\tilde{t}_2}>45$ GeV.

The superpotential involving the superfields $\hat{H}_1$, $\hat{H}_2$ and
$\hat{N}$ in the NMSSM is
\begin{equation}
W=\lambda\hat{H}_1\hat{H}_2\hat{N}-\frac{1}{3}k\hat{N}^3+
h_t \hat{Q}\hat{H}_2\hat{U}^c,
\end{equation}
where the $\hat{N}^3$ term is present to avoid a global Peccei-Quinn
$U(1)$ symmetry
corresponding to $\hat{N}\rightarrow\hat{N}{\rm e}^{i\theta}$ and
$\hat{H}_1\hat{H}_2\rightarrow \hat{H}_1\hat{H}_2{\rm e}^{-i\theta}$,
and
$\hat{Q}$ and $\hat{U}^c$ denote respectively the left-handed quark
doublet and the (anti) right handed quark singlet of the third
generation. The tree level potential is given by
\begin{eqnarray}
V_{tree}&=&V_F+V_D+V_{soft},\nonumber\\
V_F&=&\left|\lambda\right|^2\left[\left|N\right|^2\left(
\left|H_1\right|^2+\left|H_2\right|^2\right)+\left|H_1
H_2\right|^2\right]+\left|k\right|^2\left|N\right|^4\nonumber\\
&-&\left(\lambda k^{*} H_1 H_2
N^{*2}+\mbox{h.c.}\right),\nonumber\\
V_D&=&\frac{1}{8}\left(g^2+g^{\prime 2}\right)\left(
\left|H_1\right|^2-\left|H_2\right|^2\right)^2+\frac{1}{2}
g^2\left|H_{1}^{\dag}H_2\right|^2,\nonumber\\
V_{soft}&=& m_1^2\left|H_1\right|^2 +
m_2^2\left|H_2\right|^2 +
m_N^2\left|N\right|^2\nonumber\\
&-&\left(\lambda A_{\lambda}H_1 H_2 N + \mbox{h.c.}\right)-
\left(\frac{1}{3} k A_k N^3 +\mbox{h.c.}\right),
\end{eqnarray}
where
$H_1^T\equiv\left(H_1^0,H^{-}\right)$ and $H_2^T\equiv \left(H^{+},H_2^{0}
\right)$ and $g$ and $g^{\prime}$ are the gauge couplings of $SU(2)_L$
and $U(1)_Y$, respectively. Redifining the global phases of $H_2$ and $N$,
it can be shown that all
the parameters in eq. (2) can be made real, except the ratio
$r=A_{\lambda}/A$, $A_k\equiv A$.
We assume this parameter to be real \cite{alguila},
{\it i.e.} no explicit CP violation in the potential of eq. (2).

If we define
\begin{equation}
\langle H_1^{0}\rangle\equiv v_1\: {\rm e}^{i\theta_1},\:\:\:\:
\langle H_2^{0}\rangle\equiv v_2 \:{\rm e}^{i\theta_2},\:\:\:\:
\langle N\rangle\equiv x \:{\rm e}^{i\theta_3},
\end{equation}
and
\begin{equation}
3\theta_3\equiv 2\varphi_3,\:\:\:\:
\theta_1+\theta_2+\theta_3\equiv2\varphi_1,\:\:\:\:
\theta_1+\theta_2-2\theta_3\equiv2\varphi_1-2\varphi_3,
\end{equation}
we can write the most general gauge invariant (under
$SU(2)_L\otimes U(1)_Y$) potential in the vacuum
\begin{eqnarray}
\langle V\rangle&=& D_1 v_1^4+D_2 v_2^4 +D_3 v_1^2 v_2^2+D_4 v_2^2 x^2
+D_5 v_2^2 x^2 + D_6 x^4\nonumber\\
&+&D_7v_1v_2x^2\:{\rm cos}\left(2\varphi_1-2\varphi_3\right)+
D_8 m_1^2 v_1^2+ D_9 m_2^2 v_2^2 +D_{10} m_N^2 x^2\nonumber\\
&+&D_{11}v_1 v_2 x\:{\rm cos}\left(2\varphi_1\right)+
D_{12}x^3\:{\rm cos}\left(2\varphi_3\right).
\end{eqnarray}

It has been shown by Romao \cite{romao} that the minimum of this potential
at the tree
level can never be CP breaking. On the other hand, when
one-loop corrections at zero
temperature from the top-stop sector are added, CP can be spontaneously
broken \cite{bb}. However, the LEP upper bound limit on the mass
of the lightest Higgs boson particle \cite{lep}, $m_h>60$ GeV, severly
constrains the parameter space due to the general tendency
of having two neutral and
one charged Higgs boson with masses smaller than $\sim$ 110 GeV when
SCPB occurs. Moreover, the allowed area only exists for
stops heavier than $\sim$ 3 TeV, $\tan\beta=v_2/v_1\simeq 1$ and
$\lambda<0.25$ \cite{bb}.
In the following we shall
assume that CP is conserved at zero temperature so
that experimental constraints still allow large portions of the
parameter space, in particular large values of $\tan\beta$ and
$\lambda$.

The one-loop correction to the tree level potential at zero temperature
in the $\overline{DR}$ scheme of renormalization reads
\begin{equation}
V_1^{0}=\frac{1}{64\pi^2}{\rm Str}\left\{
{\cal M}^4(\phi)\left[{\rm ln}\frac{{\cal
M}^2(\phi)}{Q^2}-\frac{3}{2}\right]\right\},
\end{equation}
where ${\cal M}^2(\phi)$, with $\phi\equiv \left(H_1^0,H_2^0,N\right)$,
is the field dependent squared mass matrix, the supertrace Str properly
counts the degree of freedom, $Q$ is the renormalization point and the
$Q$ dependence in eq. (6) is compensated by that of the renormalized
parameters, so that the full effective potential is independent of $Q$
up to the next-to-leading order.

The stop squared mass
matrix in the basis $(\tilde{t}_L,\tilde{t}_R)$ is given by
\begin{equation}
{\cal M}^2_{\tilde{t}}=
\left(\begin{array}{cc}
m_{LL}^2& m_{LR}^2\\
m_{LR}^{*2}&m_{RR}^2
\end{array}\right),
\end{equation}
where
\begin{eqnarray}
m_{LL}^2&=& \tilde{m}_{q}^2+h_t^2\left|H_2^0\right|^2
+\left(\frac{g^2}{12}-\frac{g^{\prime 2}}{4}\right)\left(
\left|H_2^0\right|^2-\left|H_1^0\right|^2\right),\nonumber\\
m_{RR}^2&=& \tilde{m}_{u}^2+h_t^2\left|H_2^0\right|^2
-\frac{g^{\prime 2}}{3}\left(
\left|H_2^0\right|^2-\left|H_1^0\right|^2\right),\nonumber\\
M_{LR}^2&=&h_t\left(A_t H_2^0+\lambda N^* H_1^{*0}\right).
\end{eqnarray}

Here $A_t$ is the soft trilinear term associated to the
$\hat{Q}\hat{H}_2\hat{U}^c$ term in the superpotential. The mass matrix
is easily diagonalized by writing the sfermion stop eigenstates as
\begin{eqnarray}
\tilde{t}_1&=&-\tilde{t}_L\:\sin\theta_{\tilde{t}}+
\tilde{t}_R\:\cos\theta_{\tilde{t}},\nonumber\\
\tilde{t}_2&=&\tilde{t}_L\:\cos\theta_{\tilde{t}}+
\tilde{t}_R\:\sin\theta_{\tilde{t}},
\end{eqnarray}
whose mass eigenvalues are
\begin{equation}
m^2_{\tilde{t}_1,\tilde{t}_2}=\frac{1}{2}\left[
\left(m_{LL}^2+m_{RR}^2\right)\pm \sqrt{\left(m_{LL}^2-m_{RR}^2\right)^2+
4\left|m_{LR}^2\right|^2}\right].
\end{equation}
For large values of $A_t$, $m_{\tilde{t}_2}\ll m_{\tilde{t}_1}$ and
$\tilde{t}_2$ can be very light. Note, however, that $A_t$ is bounded
from above to
avoid dangerous color breaking minima \cite{color},
$A_{t}^{2}<3\left(\tilde{m}_{q}^2+\tilde{m}_{u}^2+m_2^2\right)$.

The one-loop correction to the tree level potential at finite
temperature reads \cite{ft}
\begin{eqnarray}
V_{1}^T&=&\frac{T^4}{2\pi^2}{\rm Str}\:J\left[\frac{{\cal
M}^2(\phi)}{T^2}\right]+\Delta V_{daisy},\nonumber\\
\Delta V_{daisy}&=&\frac{T}{12\pi}\sum_{i,bos} n_{i,bos}\left[
\left(m_{i,bos}(\phi)^2+\Pi_{i,bos}\right)^{3/2}-m_{i,bos}(\phi)^3\right],
\nonumber\\
J_{bos,fer}(y^2)&=&\int_{0}^{\infty}\:dx\:x^2\:{\rm log}\left(
1\mp{\rm e}^{-\sqrt{x^2+y^2}}\right).
\end{eqnarray}
Here the $\Pi_{i,bos}$ denotes the thermal polarization mass for each
boson with degrees of freedom $n_{i,bos}$ contributing to the Debye mass
\cite{debye}.

Defining the new parameters
\begin{eqnarray}
B_4&=& \frac{D_1+D_2}{4}-\frac{1}{4}D_3+\frac{1}{8}
\frac{D_7 D_{11}}{D_{12}},\nonumber\\
B_5&=& \frac{D_1+D_2}{4}+\frac{1}{4}D_3-\frac{1}{8}
\frac{D_7 D_{11}}{D_{12}},\nonumber\\
B_6&=&\frac{D_1-D_2}{2},\nonumber\\
B_7&=&\frac{D_4-D_5}{2},\nonumber\\
B_8&=&\frac{D_4+D_5}{2},\nonumber\\
B_9&=&D_6-\frac{1}{2}\frac{D_7D_{12}}{D_{11}},
\end{eqnarray}
whose tree level values can be inferred from eqs. (2) and (5) \cite{singlet},
it
is easy to show that CP can be spontanously broken only if the following
system of conditions is satisfied
\begin{equation}
\frac{D_{11}D_{7}}{D_{12}}>0\:\:\:{\rm plus}\:\:
\left\{ \begin{array}{c}
B_9>0,\\
B_5>0,\\
4B_5 B_9-B_8^2>0,
\end{array}\right.,
\end{equation}
which, at the tree level, does not have any solutions in the $r$ space.
The key point is the following: whenever the condition
$D_{11}D_{7}/D_{12}>0\:(\Rightarrow r<-1/3)$ assuring that the
nonvanishing phases in eq. (5) correspond to a minimum, then the
condition $B_5>0\: (\Rightarrow r>-1/3)$ is never satisfied at the tree
level. When one-loop corrections are considered, among the $B$
parameters, it is just $B_5$ which receives the largest contributions
from the top-stop sector and from the Higgs sector (at finite
temperature). It is then clear that the corrections to
the tree level potential should drive $B_5$ to positive values (for
$r<-1/3)$ to have SCPB.

A complete analysis on how to calculate the corrections to the tree
level coefficients of the most general potential, eq. (5), is given in
the Appendix of ref. \cite{singlet}.

After having made use of the minimizations equations at zero temperaure
to express the $m_1^2$, $m_2^2$ and $m_N^2$ masses in terms of the
other parameters of the potential, we define the critical temperature
$T_c$ as the value of $T$ at which the origin of the field space becomes
a saddle point \cite{singlet} which happens when one of the
$\overline{m}_i=m_{i}^2+\Pi_{i}$ ($i=1,2,N$) vanishes. The first of the
$m_{i}^2$'s to run to negative values through the renormalization group
equations is expected to be $m_{2}^2$ as a consequence of large
values of the top Yukawa coupling. As a consequence, the EWPT is
expected to occur mostly along the $H_2^0$ direction \cite{max}.

When large left-right mixing in the stop sector is present, the heaviest
stop $\tilde{t}_1$ contribution to $V_1^T$ is suppressed since
$m_{\tilde{t}_1}$ is much larger than $T$ and the values of
$\overline{m}_2^2$ is
\begin{equation}
\overline{m}_2^2=m_2^2+\frac{1}{8}\left(3g^2+g^{\prime 2}+\frac{4}{3}\lambda^2+
4h_t^2\right)T^2.
\end{equation}
In Fig. 1 we have fixed the critical temperature $T_c=150$ GeV and shown
the curve corresponding to $\overline{m}_2^2(T_c)=0$. The points in the
plane $(\lambda,m_{\tilde{t}_2})$ lying on the curve are then the points
for which $T_c=$150 GeV. Indeed, since the EWPT is known to be of the
first order, it occurs when $\overline{m}_2^2$ is still positive, {\it
i.e} at temperatures higher than $T_c$: since all the points in the
$(\lambda,m_{\tilde{t}_2})$ plane below the solid curve correspond to
$\overline{m}_2^2>0$, they correspond the region of the parameter space
where the EWPT occurs at temperatures smaller than or equal to 150 GeV.

After having calculated the corrections to the $D$ and $B$ parameters
at finite temperature with the method described in ref. \cite{singlet}
(we have included all the standard model particles as well as stops,
charginos, neutralinos, charged and neutral Higgs bosons), we have
imposed SCPB to occur inside the bubbles in the broken phase, {\it i.e.} the
satisfaction of the system in eq. (13). The results are shown in Figs 1 and 2
for different values of the parameters.

In Fig. 1 the allowed region
lies under the solid line, roughly for $\lambda<0.45$. In Fig. 2 we have
chosen $\tan\beta=1.2$ and SCPB can occur for $\lambda<0.6$. Moreover,
the lightest pseudoscalar $A^0$ should be heavier than $\sim$ 20 GeV
\cite{lepp} and the dashed line corresponds to $m_{A^0}=40$ GeV. The
lightest CP even particle $h$ has not been produced in the decay
$Z^0\rightarrow Z^{* 0}+h$ corresponding to the conservative bound
$m_h>60$ GeV \cite{lep}. These constraints impose that $\lambda$ must
lie in the range $0.1<\lambda<0.5$. We have also checked that in the
allowed regions the theory remais perturbative in the sense that
the perturbation expansion parameter $\beta\sim
g^2(T/2\pi\:\overline{m}_2)$ remains smaller than 1 \cite{debye}.

Much of the behaviour described above can be
 understood fairly well analytically when considering the largest
one-loop corrections acting mainly on $B_5$, {\it i.e} those proportional
to the top Yukawa coupling and those coming
the Higgs sector. As we
said above, for large $A_t$ the  heaviest stop $\tilde{t}_1$ contributes
only to $V_1^0$ so that its correction to $B_5$ reads
\begin{eqnarray}
(\Delta B_5)_{\tilde{t}_1}&=&\frac{(\Delta D_2)_{\tilde{t}_1}}{4}=
\frac{1}{4}\left[\frac{6}{64\pi^2}\:h_t^4\:{\rm
ln}\frac{m_{\tilde{t}_2}^2}{
\overline{Q}^2}\right.\nonumber\\
&+&\left.\frac{6}{64\pi^2}\:h_t^4\left(\frac{A_t^2}{\tilde{m}^2}
-\frac{A_t^4}{12\:\tilde{m}^4}\right)\right],
\end{eqnarray}
where we have assumed $\tilde{m}_q=\tilde{m}_u=\tilde{m}$
and neglected the gauge
couplings. $\overline{Q}$ has been properly so that the contribution
of $\tilde{t}_1$ to the minimization conditions is suppressed.

When the corrections to $B_5$ from the top and lightest stop $\tilde{t}_2$
are added, the overall {\it positive} correction $(\Delta B_5)_{t}$
from the top-stop
sector is given by
\begin{eqnarray}
(\Delta B_5)_{t}&=&\frac{(\Delta D_2)_{t}}{4}=
\frac{1}{4}\left[\frac{6}{64\pi^2}\:h_t^4\:{\rm
ln}\frac{m_{\tilde{t}_{2}}^2}{
A_f\:T^2}+\frac{6}{64\pi^2}\:h_t^4\left(\frac{A_t^2}{\tilde{m}^2}
-\frac{A_t^4}{12\:\tilde{m}^4}\right)\right.\nonumber\\
&+&\left.\frac{6}{64\pi^2}\:h_t^4\:{\rm ln}\frac{A_b}{
A_f}\right],
\end{eqnarray}
where $A_b=16\:A_f=16\:\pi^2(3/2-2\gamma_E)$, $\gamma_E$ being the Euler
constant.

As noted in ref. \cite{singlet}, the Higgs sector gives the largest
{\it negative} contribution to $B_5$
\begin{equation}
(\Delta
B_5)_{h}=-\frac{1}{8\pi}\frac{T}{\overline{m}_2}\frac{1}{16}\left(
\frac{25}{4}g^4+\frac{30}{4}g^{\prime 4}-\frac{19}{2}g^2 g^{\prime 2}
\right).
\end{equation}
Imposing now that $B_5=(B_5)_{tree}+(\Delta
B_5)_{t}+(\Delta
B_5)_{h}>0$ we find
\begin{equation}
\lambda^2<4\:\frac{(\Delta
B_5)_{t}+(\Delta
B_5)_{h}}{\left|1+3r\right|},
\end{equation}
which explains the upper bound on $\lambda$ to have SCPB inside the
bubble walls.

Here we want to make some comments. In ref.
\cite{singlet} the case $A_t\simeq 0$ and thus two nearly degenerate in
mass stops were considered. Moreover it was assumed that
$m_{\tilde{t}_1}\simeq m_{\tilde{t}_2}\simeq T$ so that both stops
contribute to $V_1^T$. In such a case it is not hard to see
that the contribution to $B_5$ from the top-stop sector is considerably
reduced and turns out to be proportional only to $h_t^4\:{\rm ln}(A_b/A_f)$.
As a consequence, the largest one-loop contribution to $B_5$ comes
from the Higgs sector and, being the latter negative, no SCPB inside the
bubbles is allowed. In this case the only possibility
is the breaking of CP inside the bubble walls, which requires some
fine-tuning among the
VEV's in the walls \cite{singlet}.
Here we have shown that, when large left-right
mixing in the stop sector is considered, SCPB can easily occur
inside the bubbles. CP violating phases are then {\it automatically}
present in the bubble walls (where, due to the strength of the
transition, several baryogenesis mechanisms are
expected to work \cite{nelson}) . We also stress that SCPB is purely driven by
plasma effects at finite temperature. Once the temperature falls down
after the end of the EWPT, the plasma corrections to the effective
potential become more and more suppressed and the CP conserving minimum
is reached. This allows to avoid a very light spectrum in the Higgs
sector which is an inevitable prediction of the Georgi Pais theorem
whenever a discrete symmetry is radiatively broken at zero temperature.
Moreover, the light stop scenario can have several intriguing
experimental consequences which could show up in the next generation
of accelerators.

\newpage
%

\newpage
\noindent{\bf Figure Caption}
\begin{itemize}
\item[{\bf Fig. 1)}]{The allowed region for having SCPB in the plane
$(\lambda,m_{\tilde{t}_2})$ lies under the solid line. Here
$\tilde{m}=400$ GeV, $A=50$ GeV, $x=250$ GeV, $T=150$ GeV,
$m_t=174$ GeV, $r=-1.2$, $\tan\beta=10$ and $k=0.6$. The curve
$\overline{m}_2=0$ is also indicated.}
\end{itemize}
\begin{itemize}
\item[{\bf Fig. 2)}]{The allowed region for having SCPB in the plane
$(\lambda,m_{\tilde{t}_2})$ lies under the solid line. Here
$\tilde{m}=400$ GeV, $A=50$ GeV, $x=250$ GeV, $T=150$ GeV,
$m_t=174$ GeV, $r=-0.8$, $\tan\beta=1.2$ and $k=0.5$. The curves
$m_A=40$ GeV and $m_h=60$ GeV are also indicated.}
\end{itemize}
\end{document}